\documentclass[twocolumn]{aastex62}
\usepackage{amsmath}
\usepackage{url}
\received{}
\revised{}
\accepted{}

\shorttitle{}
\shortauthors{}

\begin{document}

\title{\Large Observing the redshifted 21 cm signal around a bright QSO at $z\sim 10$}

\correspondingauthor{Qing-Bo Ma}
\email{maqb@gznu.edu.cn}

\author[0000-0001-9493-4565]{Qing-Bo Ma}
\affil{School of Physics and Electronic Science, Guizhou Normal University, Guiyang 550001, PR China}
\affil{Guizhou Provincial Key Laboratory of Radio Astronomy and Data Processing, \\
Guizhou Normal University, Guiyang 550001, PR China}

\author{Benedetta Ciardi}
\affiliation{Max-Planck-Institut f\"ur Astrophysik, Karl-Schwarzschild-Stra\ss e 1, D-85748 Garching bei M\"unchen, Germany}

\author{Koki Kakiichi}
\affiliation{Department of Physics and Astronomy, University College London, London WC1E 6BT, UK}

\author{Saleem Zaroubi}
\affil{Kapteyn Astronomical Institute, University of Groningen, PO Box 800, NL-9700 AV Groningen, the Netherlands}
\affil{Department of Natural Sciences, Open University of Israel, 1 University Road, PO Box 808, Ra'anana 4353701, Israel}
\affil{Department of Physics, The Technion, Haifa 32000, Israel}

\author{Qi-Jun Zhi}
\affil{School of Physics and Electronic Science, Guizhou Normal University, Guiyang 550001, PR China}
\affil{Guizhou Provincial Key Laboratory of Radio Astronomy and Data Processing, \\
Guizhou Normal University, Guiyang 550001, PR China}

\author{Philipp Busch}
\affiliation{Max-Planck-Institut f\"ur Astrophysik, Karl-Schwarzschild-Stra\ss e 1, D-85748 Garching bei M\"unchen, Germany}



\begin{abstract}
We use hydrodynamics and radiative transfer simulations to study the 21~cm signal around a bright QSO at $z \sim 10$.
Due to its powerful UV and X-ray radiation, the QSO quickly increases the extent of the fully ionized bubble produced by the pre-existing stellar type sources, in addition to partially ionize and heat the surrounding gas.
As expected, a longer QSO lifetime, $t_{\rm QSO}$, results in a 21~cm signal in emission located at increasingly larger angular radii, $\theta$, and covering a wider range of $\theta$. Similar features can be obtained with a higher galactic emissivity efficiency, $f_{\rm UV}$, so that determining the origin of a large ionized bubble (i.e. QSO vs stars) is not straightforward. Such degeneracy could be reduced by taking advantage of the finite light traveltime effect, which is expected to affect an HII region produced by a QSO differently from one created by stellar type sources.
From an observational point of view, we find that the 21 cm signal around a QSO at various $t_{\rm QSO}$ could be detected by SKA1-low with a high signal-noise ratio (S/N). As a reference,  for $t_{\rm QSO} = 10\,\rm Myr$, a S/N $\sim 8$ is expected assuming that no pre-heating of the IGM has taken place due to high-$z$ energetic sources, while it can reach value above 10 in case of pre-heating. Observations of the 21~cm signal from the environment of a high-$z$ bright QSO could then be used to set constraints on its lifetime, as well as to reduce the degeneracy between $f_{\rm UV}$ and $t_{\rm QSO}$.
\end{abstract}


\section{Introduction}
\label{sec:intro}
After the formation of the first structures in the Universe (see \citealt{Ciardi2005,Bromm2013,Dayal2018} for reviews), their UV and X-ray radiation starts to propagate into the surrounding neutral gas and initiates the reionization process, which should be complete by $z \simeq 6$, as suggested by the Gunn-Peterson through in QSO spectra (\citealt{Fan2006,Fan2006b}, but see e.g. \citealt{Kulkarni2019} for a recent discussion on the possibility of having a later reionization).
This era is referred to as epoch of reionization (EoR).
Cosmic microwave background experiments, most recently the Planck telescope, measured a Thomson scattering optical depth $\tau = 0.0544 \pm 0.007$ \citep{Planck2018}, suggesting a global ionization fraction of $0.5$ at redshift $7.68\pm0.79$.
However, more details on the EoR are expected from 21~cm experiments, such as the Low-Frequency Array (LOFAR)\footnote{http://www.lofar.org/}, the Murchison Widefield Array (MWA)\footnote{http://www.mwatelescope.org/}, the Hydrogen Epoch of Reionization Array (HERA)\footnote{https://reionization.org/}, and the upcoming Square Kilometre Array (SKA)\footnote{https://www.skatelescope.org/}.

Although the first sources (e.g. metal-free stars and mini-QSOs) are predicted to have bright radiation which initiates and contributes to the reionization process \citep{Chen2008, Ghara2016}, they cannot provide the full photon budget required to complete reionization \citep{Choudhury2006, Trac2007}.
The bulk of H-ionizing photons is instead produced by subsequent stellar generations, together with minor contributions from more energetic sources such as X-ray binaries, QSOs, and shock heated interstellar medium \citep{Dijkstra2004, Eide2018}.
While any single one of these sources is typically too faint to be detected (but refer also to \citealt{Ghara2016,Ghara2017} for a more optimistic view), their integral contribution could be measured by 21~cm experiments, e.g. in terms of 21~cm power spectra \cite[e.g. ][]{Madau1997,Geil2009, Christian2013, Patil2014, Seiler2018, Ross2019}, 21~cm bispectra \cite[e.g. ][]{Shimabukuro2017, Watkinson2019}, cross correlations of 21~cm with other observations \cite[e.g. ][]{Vrbanec2016, Ma2018b, Moriwaki2019}.
An exception is constituted by sources as bright as a QSO that would carve large ionized regions \citep{Alvarez2007, Feng2013, Kakiichi2017}, which could be resolved by 21 cm telescopes \citep{Geil2008, Majumdar2011, Datta2012, Datta2016, Bolgar2018}, and possibly used to set constraints on the ionizing photon rate and/or lifetimes of the quasar \citep{Wyithe2004, Datta2012},  as well as to distinguish a reionization history dominated by QSOs from one dominated by stellar type sources  \cite[e.g. ][]{Hassan2019}.
These bright sources could also be directly detected by optical/near-infrared telescopes such as the Thirty Meter Telescope (TMT), the Euclid telescope and the James Webb Space Telescope (JWST).
Combining information from such different observations, is expected to offer the possibility of a more thourough investigation of the ionizing and heating process of the IGM, as well as of the source properties during the EoR.

In this paper, we study the 21~cm signal associated to a bright QSO at $z \sim 10$ using the simulations described in \citet[][hereafter K17]{Kakiichi2017}, and the detectability of such signal with LOFAR and SKA. Incidentally, this is within the range of redshift of the LOFAR peak performance, i.e. $z\sim 8.5-10.5$ (\citealt{Patil_etal_2017}).
As mentioned earlier, similar works appeared in the literature, employing both analytic models \cite[e.g.][]{Wyithe2004, Majumdar2011} and radiative transfer simulations \cite[e.g.][]{Datta2012, Bolgar2018}.
Building on these, K17 has run a suite of 3D cosmological N-body/hydrodynamical and multifrequency radiative transfer simulations to model the QSO environment (i.e. the surrounding galaxies and the intergalactic medium), the spectral shape of the radiation emitted by the QSO as well as the galaxies, and the propagation of such radiation with a consistent treatment of UV and X-ray photons, and secondary ionization.
Here we will present an investigation of the 21~cm signal around the high-$z$ bright QSO investigated in K17.

The cosmological parameters adopted are from the  Wilkinson Microwave Anisotropy Probe 9 year result \citep{Hinshaw2013}: $\Omega_{\Lambda}= 0.74$, $\Omega_{m} = 0.26$, $\Omega_{b} = 0.0463$, $h = 0.72$, $n_{s} = 0.95$ and $\sigma_{8} = 0.85$.
The layout of the paper is as follows.
Section~\ref{sec:simu} describes the simulations adopted.
The expected 21~cm signal and signal-noise ratio are in Section~\ref{sec:res}.
The conclusions and discussions are in Section~\ref{sec:con}.

\section{Simulations}
\label{sec:simu}
For this study we adopt the simulations discussed in K17.
Here we outline the features relevant to this work, while we refer the reader to the original paper for more details.

As a first step, the IGM and galaxies were modeled by running until $z=10$ the smoothed particle hydrodynamics code GADGET-3 \citep{Springel2005} with $2\times 512^{3}$ dark matter and gas particles.
The simulation box has a length of $50\, h^{-1}\, \rm cMpc$ and it is by design centered on the largest halo, with a mass $M_{\rm halo}=1.34\times 10^{10} \,h^{-1}\,\rm M_{\odot}$ at $z=10$.

Then, the gas density and temperature were mapped onto a Cartesian grid with $256^{3}$ cells, to be used as input for the radiative transfer (RT) code CRASH \citep{Ciardi2001,Maselli2003,Maselli2009, Graziani2013, Graziani2018}, which models the gas temperature and ionization (of both its H and He components) by following the propagation of ionizing photons in the frequency range 13.6~eV-2~keV.
The stellar type sources hosted by galaxies were turned on at $z=15$ with an ionizing photon rate linearly related to the halo mass:
\begin{equation}
    \dot{N}_{\rm ion}^{\rm GAL}(M_{\rm halo})=\dot{n}_{\rm ion}(z) \frac{M_{\rm halo}} {\bar{\rho}_{\rm halo}},
\end{equation}
where $\bar{\rho}_{\rm halo}=\sum^{N_{s}}_{j=1} M_{\rm halo}^{j}/V_{\rm box}$,
\begin{equation}
    \dot{n}_{\rm ion}(z)=10^{50.89}\lambda(z)\frac{\alpha_{\rm b}+3}{2\alpha}f_{\rm UV},
\end{equation}
$V_{\rm box}$ is the comoving volume of the simulation box, $N_{s}$ denotes the total number of galaxies from the simulation, $\alpha$ ($\alpha_{\rm b}$) is the power-law spectral index of the sources (the ionizing background),
$\lambda(z)$ describes the redshift dependence of the star formation rate density, and $f_{\rm UV}$ is the emissivity efficiency.
$\alpha$ and $\alpha_{\rm b}$ are assumed to be 3.
The adopted expression for $\lambda(z) =\frac{\xi e^{\zeta(z-9)}}{\xi-\zeta+\zeta e^{\xi(z-9)}}$ with $\xi = 14/15$ and $\zeta = 2/3$ is from \cite{Bolton2007} and \cite{Ciardi2012}.
Considering the uncertainty in galactic luminosities and escape fraction during the EoR, three $f_{\rm UV}$ values have been adopted, i.e. $f_{\rm UV}=0.05$, 0.1 and 0.2, which are in the range allowed by current experiments.
For each of the three simulations, another was run in which a QSO located in the most massive halo in the center of the box, which was turned on at $z=10$ for a time
$t_{\rm QSO} = 10$~Myr, i.e. until $z = 9.85$.
Its ionizing photon production rate was modeled by re-scaling the properties of ULAS J1120+0641 \citep{Bolton2011,Mortlock2011} at $z = 7.085$ to $z=10$, i.e. $\dot{N}_{\rm ion}^{\rm QSO} = 1.36 \times 10^{56}\, \rm photons \, s^{-1}$.
We assumed its spectrum to be a power law $L_{\nu}^{\rm QSO}(\nu)\propto\nu^{-1.5}$ and its radiation to be spherically symmetric.

A total of six simulations are included in this paper, three with and three without the QSO contribution.
In K17 they are named GAL$\_$R+QSO$\_$UVXsec (our fiducial model, which includes the effect of the QSO and adopts  $f_{\rm UV}=0.1$), GAL$\_$0.5R+QSO$\_$UVXsec, GAL$\_$2R+QSO$\_$UVXsec, GAL$\_$R, GAL$\_$0.5R and GAL$\_$2R.

\begin{figure}
    \centering
    \includegraphics[height=\linewidth,angle=90]{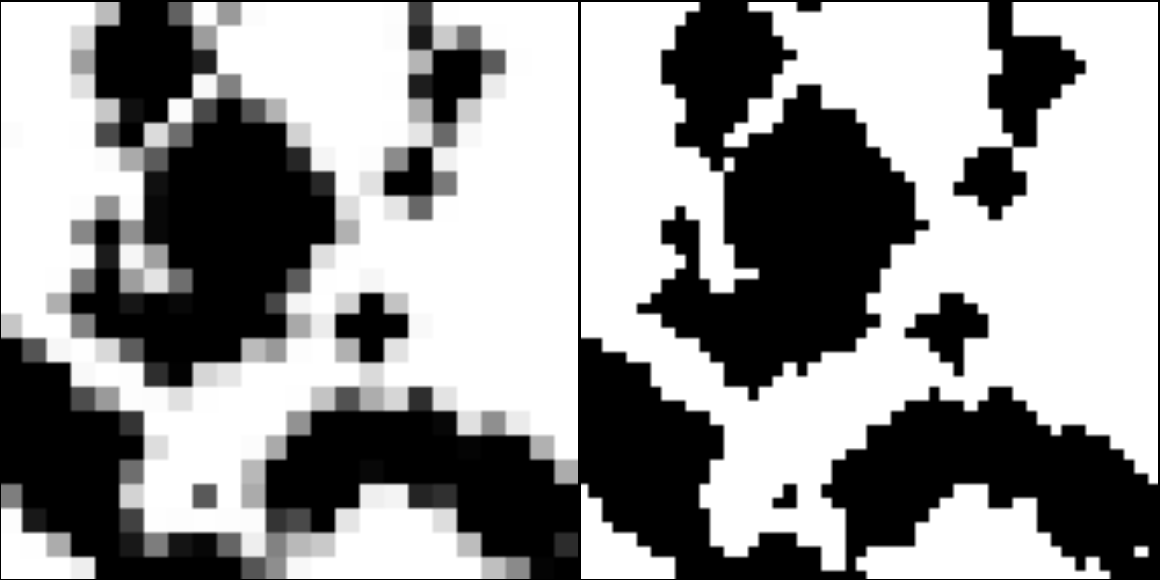}
    \caption{Example slice of an ionization field before (left panel) and after (right panel) our correction procedure as described in \autoref{sec:simu}. The gray-scale map gives the ionization field in linear units with black fully ionized and white completely neutral. The splitting of one layer of cells into 8 sub-cells per cell results in two layers of sub-cells of which we are only showing one in this example.}
    \label{fig:ref_ex}
\end{figure}

Finally, large scale simulations as those considered here are affected by resolution contamination in the cells containing the ionization front (I-front). When stars are the only sources of radiation, the I-front is expected to be very sharp due to the small mean free path of the UV photons and thus it can not be resolved. As a consequence, the cells which contain it  appear as partially ionized and warm \citep{Ross2017}, while in reality part of the gas in the cell should be neutral and cold, and part fully ionized and hot. As this issue can affect the estimate of the 21~cm signal, we correct for it using a post-processing technique which exploits the knowledge about the dominance of stellar emission in fully ionized regions and the negligible width (compared to the cell dimension) of the corresponding I-front.

More specifically, we divide each partially ionized cell in $n_{sub}$ (in this work 8; we have checked that a value of 64 gives the same results) sub-cells that are either fully ionized or completely neutral\footnote{Note that, for consistency in the analysis, also fully ionized/neutral cells can be equally refined, but the sub-cells in this case have the same physical properties of the original cells.}.
Assigning to all sub-cells the same density of the original cell, the number of fully ionized sub-cells, $n_{ion}$, is:
\begin{equation}
 n_{ion} = \mathrm{round}\left(x_{\mathrm{HII}}\cdot n_{sub} \right),
\end{equation}
where $x_{\mathrm{HII}}$ is the ionization fraction of the original cell. This procedure also assures that the average ionization fraction of the sub-cells closely matches that of the original cell.

To decide which sub-cells will be fully ionized or fully neutral, we minimize their distance to fully ionized cells by calculating the Euclidean distance transform \citep{Rosenfeld1966} and choosing the $n_{ion}$ sub-cells with the smallest distance.
We iterate this process until it converges, usually after two to four iterations.
This procedure leaves us with compact arrangements of sub-cells that smoothly trace the original borders.
The temperature in the ionized sub-cells is set to the temperature of the parent cell, while in the neutral ones it is set to the value from the original hydrodynamical simulation.
Note that although this method is likely to underestimate the temperature of the ionized sub-cells, as they do not contribute to the emission of the 21~cm signal, our assumption does not affect the final results.

An example of the outcome in terms of the ionization field produced by stellar type sources is given in \autoref{fig:ref_ex}, where we show a slice of the field before and after the correction procedure described above. We clearly see that the partially ionized cells (gray cells in the left panel) are now split into fully neutral/ionized sub-cells.

The procedure is slightly different when radiation beyond the UV range is present, as harder photons indeed lead to extended, partially ionized, warm regions.
In these cases the first layer of partially ionized cells outside fully ionized regions is corrected by setting their ionization level and temperature to those of the closest cell not in direct contact with fully ionized regions and therefore only affected by the long mean free path radiation.

\section{Results}
\label{sec:res}
In this section we will evaluate the 21~cm signal expected from the QSO environment, as well as discuss its detectability.

\subsection{Expected 21~cm signal}
The brightness temperature of the 21~cm signal can be expressed as \citep{Furlanetto2006}:
\begin{equation}
\label{eq:t21cm}
\begin{split}
\delta T_{\rm 21cm}=27 \,{\rm mK} \times x_{\rm HI} (1+\delta) \left (1-\frac{T_{\rm CMB}}{T_{S}}\right )\\ 
\times \left (\frac{1+z}{10}\frac{0.15}{\Omega_{m}h^2} \right )^{1/2} \left (\frac{\Omega_{b}h^2}{0.023} \right ),
\end{split}
\end{equation}
where $x_{\rm HI}$ is the fraction of neutral hydrogen, $\delta$ is the gas overdensity, $T_{\rm CMB}$ is the CMB temperature, and $T_{S}$ is the spin temperature, for which we make the reasonable assumption to be fully coupled to the gas temperature $T$ at the redshifts of interest.
In the following, we will consider two cases: one in which the only sources of heating are those included in our simulations, i.e. $T_S=T$, and a second one in which we assume that the IGM is preheated in the early stages of cosmic reionization by more energetic sources like nuclear black holes, X-ray binaries and/or shock heated interstellar medium (see e.g. \citealt{Eide2018}), i.e. $T_S \gg T_{\rm CMB}$.
We have tested how peculiar velocities affect our results using the MM-RRM (Mesh-to-Mesh Real-to-Redshift-Space-Mapping) method described in \cite{Mao2012} to include redshift space distortions. As we find that this correction has an effect at a percentage level, we ignore it in the remainder of the paper to simplify the discussion.

\begin{figure}
    \centering
    \includegraphics[width=\linewidth]{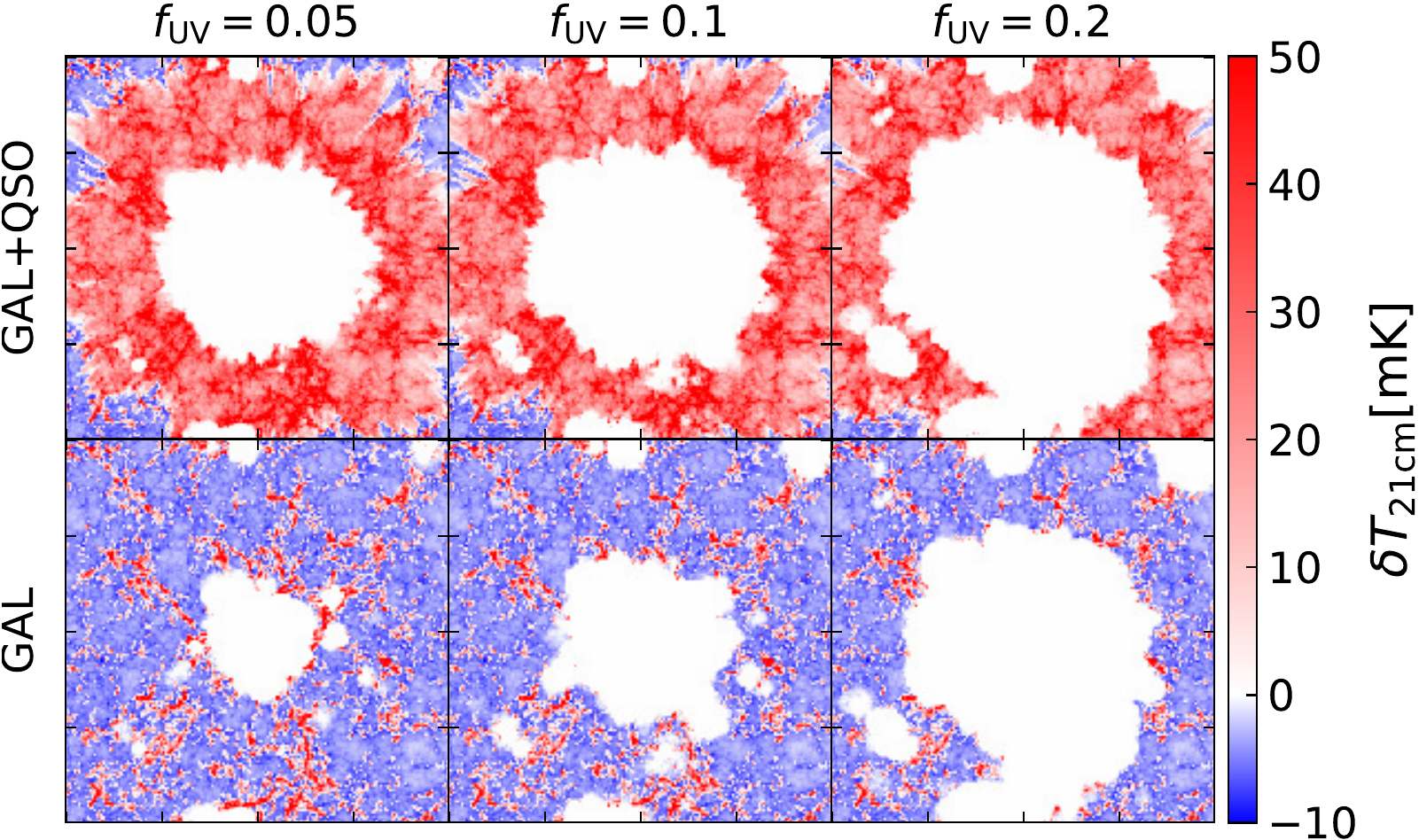}
    \includegraphics[width=\linewidth]{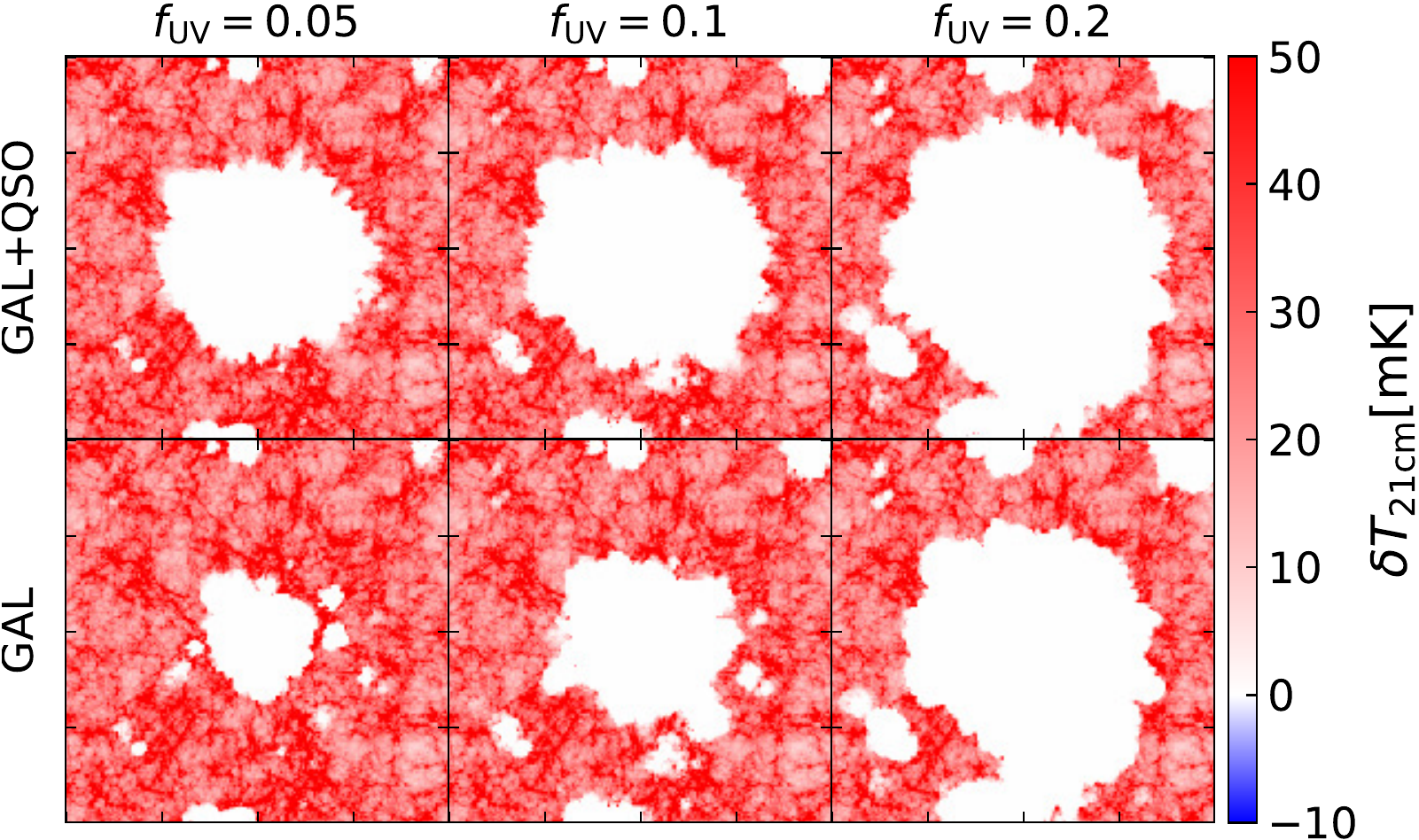}
    \caption{{\it Top figure:} Central slices of 21~cm signal from the simulations with (upper panels) and without (lower panels) QSO in the case of $T_{S} = T$. From left to right the columns refer to $f_{\rm UV} = 0.05$, 0.1 and 0.2. The length of each axis is $50\,h^{-1}\,\rm cMpc$. The maps refer to $t_{\rm QSO}=10$~Myr ($z = 9.85$). {\it Bottom figure:} Same as the top except that $T_{S} \gg T_{\rm CMB}$.}
    \label{fig:s_21cm_ts}
\end{figure}
Fig.~\ref{fig:s_21cm_ts} shows the central slices of the 21~cm signal around the QSO for $t_{\rm QSO}=10$~Myr (corresponding to $z = 9.85$ and an observed frequency of $\nu = 130.9 \, \rm MHz$), together with the results of simulations without QSO.
As already discussed in K17, the presence of a bright QSO increases the extent of the fully ionized region, although its effect becomes less evident with increasing $f_{\rm UV}$ (i.e. galactic luminosity).
On the other hand, because of the longer mean free path of the more energetic photons emitted (which extend into the soft X-ray regime), the presence of a QSO is characterized by a ring of warm and partially ionized gas, which is absent when only stellar type sources are considered.
In the case of $T_S=T$, this results in a strong positive 21~cm signal just outside the fully ionized region, while it is much smaller in simulations without QSO.
Note that in the later case the positive 21~cm signal is due to the shockwave heating modeled in the hydrodynamic simulation.
If we are instead in a regime in which $T_{S} \gg T_{\rm CMB}$, the 21~cm signal is always positive.
Compared to simulations with only stellar sources, the maps with a QSO have no obvious feature indicating its presence, except for the slightly larger size of the ionized bubble.

\begin{figure}
    \centering
    \includegraphics[width=0.99\linewidth]{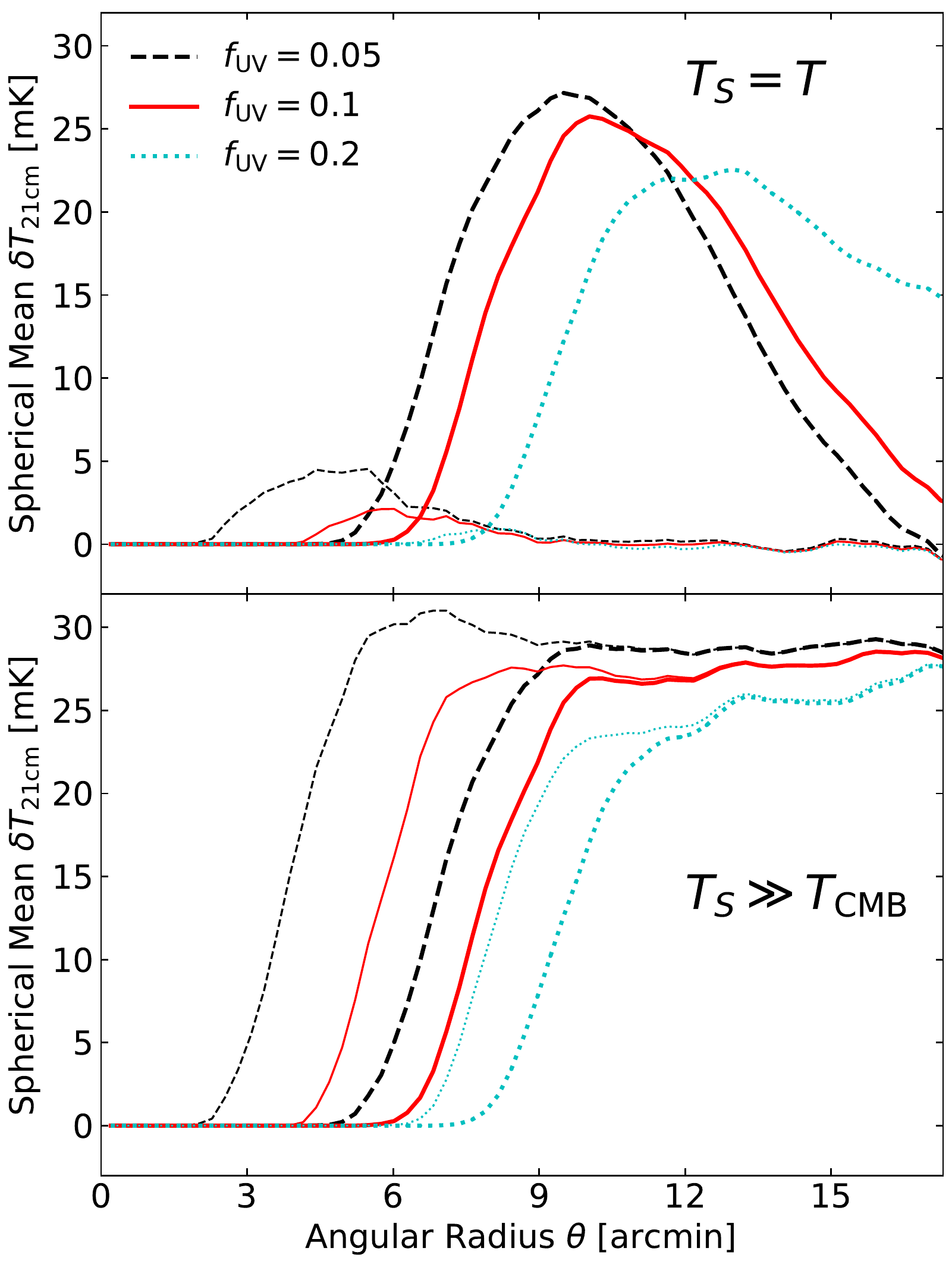}
        \caption{Spherically averaged 21~cm signal from the simulations with (thick lines) and without (thin lines) QSO. The lines refer to $f_{\rm UV} = 0.05$ (black dashed), 0.1 (red solid) and 0.2 (cyan dotted). The top and bottom panels refer to the case with $T_{S} = T$ and $T_{S} \gg T_{\rm CMB}$, respectively. The profiles are shown for $t_{\rm QSO}=10$~Myr ($z=9.85$).}
    \label{fig:s_21cm_mean_ang}
\end{figure}
Fig.~\ref{fig:s_21cm_mean_ang} shows the spherically averaged 21 cm signal for $t_{\rm QSO}=10$~Myr as a function of the angular radius $\theta$ with the simulation box center as zero point, i.e. the 21 cm signal distribution around the central halo hosting the QSO.
The top panel presents the results assuming $T_{S} = T$ for the simulations including only galaxies (thin lines) and both galaxies and the QSO (thick lines), for three different values of $f_{\rm UV}$.
With only galaxies, the 21~cm signal is always much weaker ($\delta T_{\rm 21cm} < 5\,\rm mK$) than with the inclusion of the QSO.
As mentioned earlier, the signal in emission is due to shock heated gas and thus is roughly proportional to the gas density. As a consequence, the amplitude of the signal outside the ionized bubble decreases with increasing radius. For the same reason, the simulation with higher $f_{\rm UV}$ produces a larger ionized bubble and thus a lower 21~cm amplitude. In the simulation with both galaxies and QSO,
the presence of the QSO manifests itself as a strong emission peak corresponding to the partially ionized ring observed in the previous figure,
with the highest value of $\delta T_{\rm 21cm}\sim 27 \,\rm mK$ reached at $\theta \sim 9.5\,\rm arcmin$ in the simulation with $f_{\rm UV} = 0.05$.
The UV emitting efficiency $f_{\rm UV}$ not only relates to the size of ionized bubbles, but also affects the amplitude and distribution of the average 21~cm signal. For example,
if $f_{\rm UV}$ is increased to 0.1 and 0.2, the location of the peak shifts to larger angular radii, i.e. to $\theta \sim 10 \,\rm arcmin$ and $13\,\rm arcmin$, respectively.
Meanwhile, the intensity of the peak decreases to $\delta T_{\rm 21cm}\sim 22.5 \,\rm mK$ and $\sim 20 \,\rm mK$, since a higher $f_{\rm UV}$ leads to a lower mean $n_{\rm HI}$ in the partially ionized ring.

The bottom panel presents the same results with $T_{S} \gg T_{\rm CMB}$.
In this case, the amplitude of the 21~cm signal is proportional to $x_{\rm HI}(1+\delta)$,
thus the signal is always positive, increasing rapidly outside the fully ionized region and reaching a value of $\delta T_{\rm 21cm} \sim 27 \,\rm mK$ independently from the presence of the QSO.
The only effect of the QSO is that, by increasing the size of the ionized bubble, it pushes the emission to larger radii.
While this is more evident for lower values of $f_{\rm UV}$, for which the profiles are more dissimilar, they nevertheless do not present any characteristic that would point towards the presence of a QSO.
In fact, results from simulations with only but more luminous stellar sources could be very similar to those with a QSO and lower $f_{\rm UV}$ (e.g. the thin cyan dotted and thick red solid lines in the bottom panel of Fig.~\ref{fig:s_21cm_mean_ang}),
while larger $f_{\rm UV}$ usually give lower values of $\delta T_{\rm 21cm}$ because of the lower $n_{\rm HI}$, similarly to what observed for $T_S=T$.

\begin{figure}
    \centering
    \includegraphics[width=0.99\linewidth]{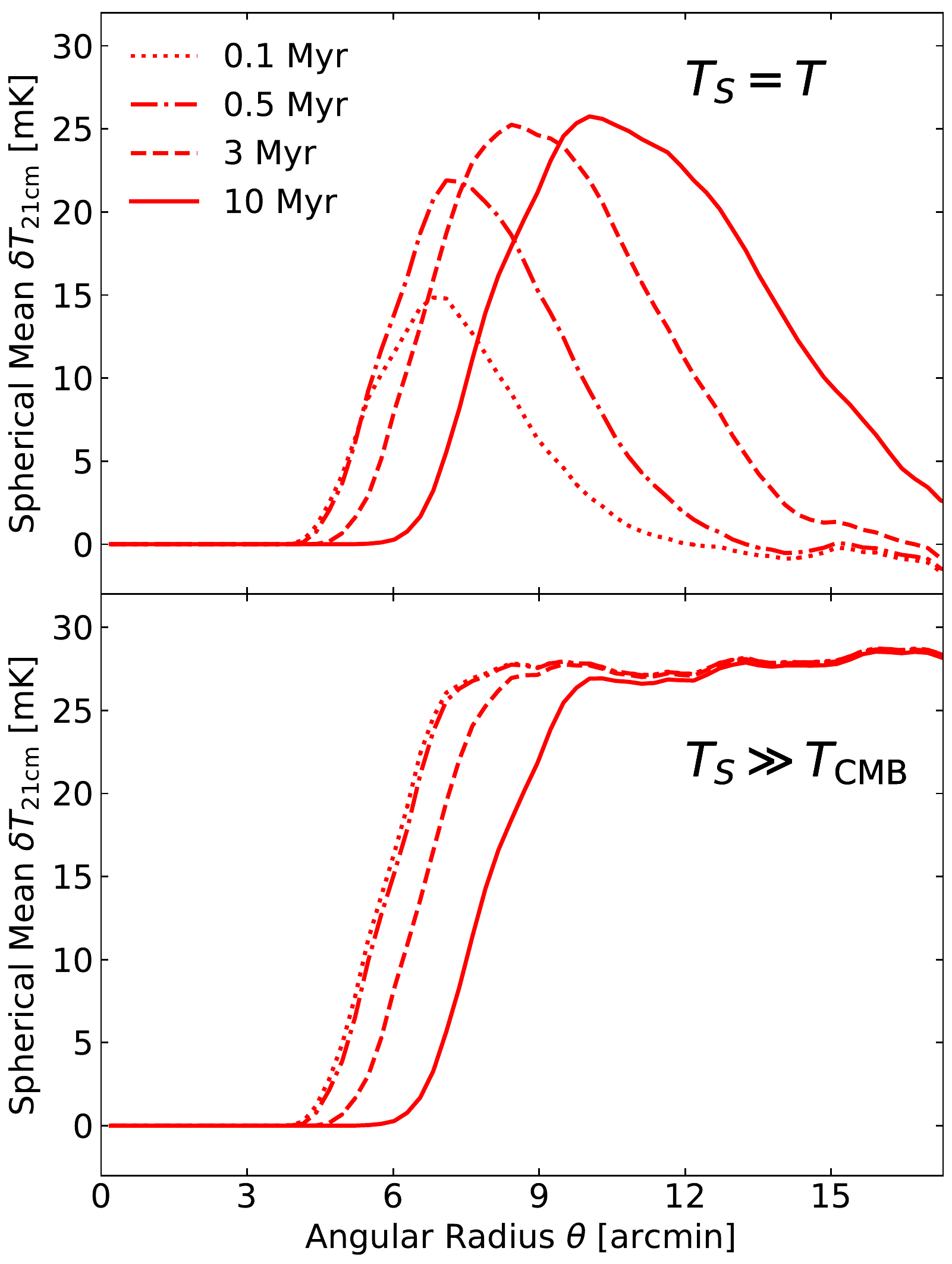}
    \caption{Spherically averaged 21 cm signal from our fiducial model with $t_{\rm QSO} = 0.1\,\rm Myr$ (dotted), $0.5\,\rm Myr$ (dash-dotted), $3\,\rm Myr$ (dashed), and $10\,\rm Myr$ (solid). The top and bottom panels refer to the case with $T_{S} = T$ and $T_{S} \gg T_{\rm CMB}$, respectively.
    }
    \label{fig:s_21cm_mean_ang_vz}
\end{figure}
Fig.~\ref{fig:s_21cm_mean_ang_vz} shows the spherically averaged 21~cm signal in the fiducial model for various ages of the QSO.
The 21~cm signal with $T_{S} = T$ shows obvious evolution in a short time, with a peak of the emission moving quickly away from the central source.
Due to its high energy radiation, the QSO can quickly (i.e. less than 0.1~Myr) heat the surrounding gas well above $T_{\rm CMB}$ and induce a signal in emission.
As $t_{\rm QSO}$ increases, both the fully ionized bubble and the range of angular radii ($\Delta \theta$) of positive 21~cm signal rapidly grow in size (e.g. $\Delta \theta$ is $ \sim 8\,\rm arcmin$ at $t_{\rm QSO}=3\,\rm Myr$ while $\sim 12\,\rm arcmin$ at $t_{\rm QSO}=10\,\rm Myr$).
The enlarging ionized bubble is also clearly visible in 21~cm signal with $T_{S} \gg T_{\rm CMB}$.
In this case,  the signal converges at large distances from the central source.
Our results, for both assumptions on $T_S$, confirm previous suggestions (see e.g. \citealt{Datta2012}) that the rapid evolution
 of the 21~cm signal caused by a QSO could be used to constrain its age.

\cite{Majumdar2011} suggested that the finite light traveltime (FLTT) effect is expected to affect the observed 21~cm signal from an HII region produced by a QSO much more significantly than one by galaxies.
More specifically, the observed 21 cm signal around a QSO is expected to be very anisotropic along  the line of sight (LOS) \citep{Majumdar2011, Zawada2014}, while that from only stellar sources would be more spherical.
As a consequence, measuring the 21~cm signal along the LOS could be used to disentangle its origin.
Additionally, as already discussed, a 21~cm signal similar to the one produced by a QSO with a given $t_{\rm QSO}$ could be obtained also by galaxies alone with an appropriate value of $f_{\rm UV}$ (see Fig.~\ref{fig:s_21cm_mean_ang}).
On the other hand, as the evolution in the case of a QSO is much more rapid, observations along the LOS to the QSO should reveal a signal different than in the case of galaxies only, and thus might be used to break the degeneracy between $t_{\rm QSO}$ and $f_{\rm UV}$.
\begin{figure}
    \centering
    \includegraphics[width=0.99\linewidth]{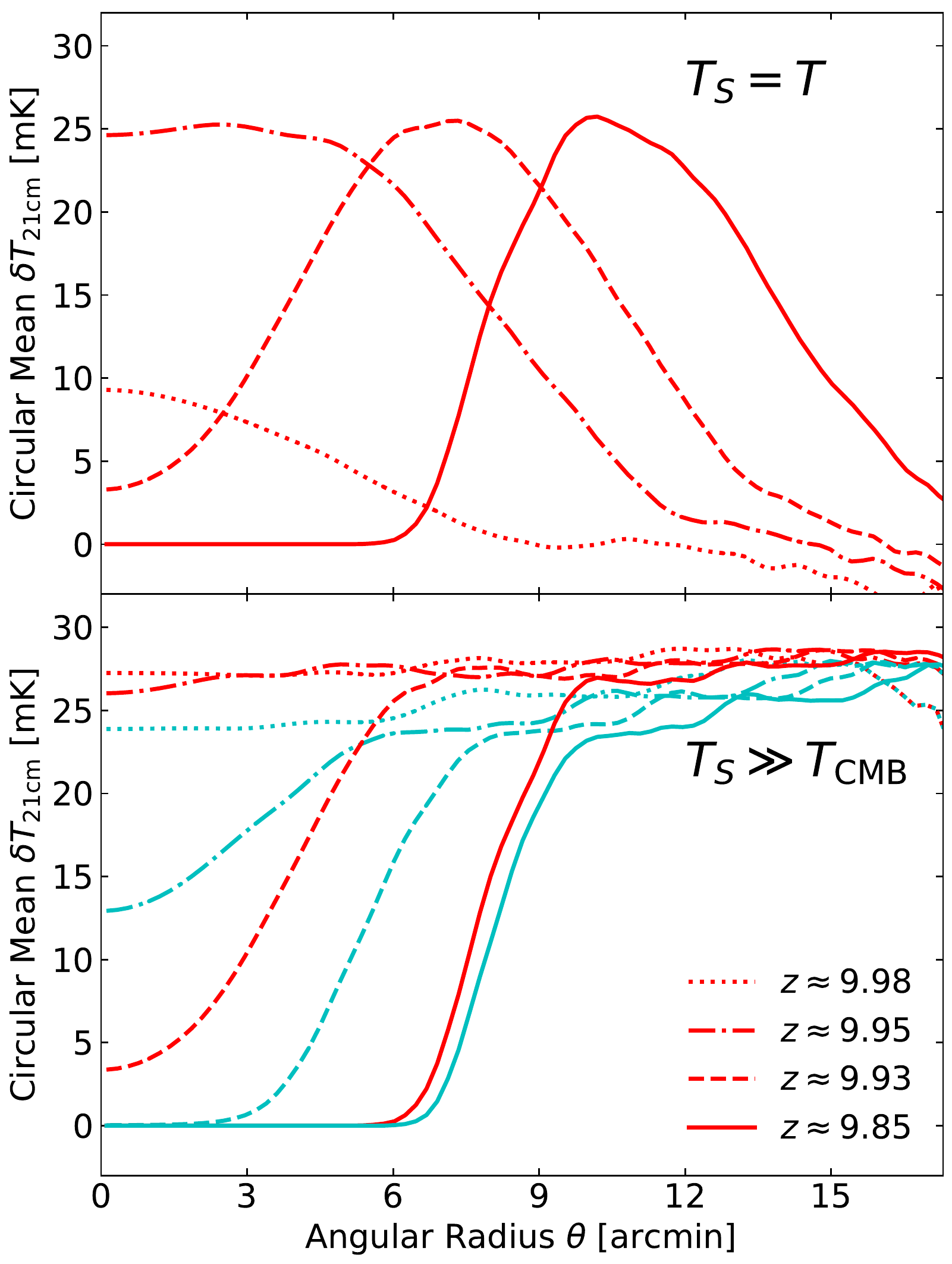}
    \caption{Circularly averaged 21 cm signal in slices perpendicular to the line of sight to the QSO at $z=9.85$ in our fiducial model. The slices correspond to $z= 9.98$ (dotted), 9.95 (dash-dotted), 9.93 (dashed)  and  9.85 (solid lines), i.e. $\nu= 129.3$, 129.7, 130.0 and 130.9~MHz.
    The top and bottom panels refer to the case with $T_{S} = T$ and $T_{S} \gg T_{\rm CMB}$, respectively.
    The bottom panel also includes the results from the model without QSO but $f_{\rm UV} = 0.2$ (thin cyan lines).
    }
    \label{fig:s_21cm_mean_ang_los}
\end{figure}
To test how the 21 cm signal behaves for our QSO model by considering the FLTT effect,
Fig.~\ref{fig:s_21cm_mean_ang_los} shows the circularly averaged
21 cm signal of slices perpendicular to the LOS to the QSO of $t_{\rm QSO}=10$~Myr ($z = 9.85$) in our fiducial model.
The slices refer to  $z= 9.98$, 9.95 and 9.93 (i.e. $\nu= 129.3$, 129.7 and 130.0~MHz) that have comoving distances of 30, 22.8 and 16.6~Mpc to the QSO\footnote{Note that, since our RT simulations do not include the effect of finite speed of ionizing photons, here we only consider the FLTT effect of the 21 cm photons that travel from the neutral hydrogen to the observer.}.
With $T_{S} =T$, the 21~cm positive signal extends to $z = 9.98$, although at this redshift it is much weaker than that at lower redshift.
While the characteristic ring of 21~cm emission associated to the presence of a large bubble carved by the QSO is clearly visible until $z=9.93$, with a peak at $\theta \sim 7.3 \, \rm arcmin$, at $z>9.95$ it disappears.
With $T_{S} \gg T_{\rm CMB}$, only the 21 cm signal at $z \lesssim 9.93$ shows the features of an ionized bubble, while the signal at higher $z$ resembles very closely that of the neutral gas medium.

To investigate if indeed the FLTT effect can reduce the degeneracy between $f_{\rm UV}$ and $t_{\rm QSO}$ when $T_{S} \gg T_{\rm CMB}$, in the bottom panel of Fig.~\ref{fig:s_21cm_mean_ang_los} we also present the results from the model without QSO but $f_{\rm UV} = 0.2$.
As mentioned earlier, this case has a 21 cm signal at $z= 9.85$ similar to that of the fiducial model, but at $z= 9.93$ and 9.95 it displays an obviously lower signal at $\theta < 6 \,\rm arcmin$.
Further away from the central source, e.g. at $z= 9.98$, the 21 cm signal is equivalent to one from an almost fully neutral IGM, i.e. an almost constant value with increasing $\theta$, both with and without the QSO.

\subsection{Detectability of the 21~cm signal}
In this section we investigate the feasibility of observing the 21~cm signal around the high-$z$ QSO with radio interferometers like LOFAR and SKA1-low.
The flux density noise of an interferometer can be expressed as \citep{Wilson2009}:
\begin{equation}
    \sigma_{\rm S} =\frac{2 k_{\rm B} T_{\rm sys}}{\epsilon A_{\rm eff} \sqrt{N_{\rm st} (N_{\rm st}-1) B t_{\rm int}}},
\end{equation}
where $k_{\rm B}$ is the Boltzmann constant, $T_{\rm sys}$ is the system temperature, $\epsilon$ is the efficiency factor, $A_{\rm eff}$ is the effective collecting area, $N_{\rm st}$ is the number of stations, $B$ is the frequency bandwidth and $t_{\rm int}$ is the integration time.
$S = \epsilon A_{\rm eff} / T_{\rm sys}$ denotes the sensitivity of one station.
Using the Rayleigh-Jeans relation $ \sigma_{\rm S} = 2 k_{\rm B} \Delta T_{b} \Omega_{\rm beam} \lambda^{-2}$, the brightness temperature of the instrumental noise can then be expressed as:
\begin{equation}
    \delta T_{\rm N} =  \frac{\lambda^{2}}{S \Omega_{\rm beam} \sqrt{N_{\rm st} (N_{\rm st}-1) B t_{\rm int}}},
\end{equation}
where $\Omega_{\rm beam} = 1.133\vartheta^{2}$ and $\vartheta$ is the angular resolution.
Note that here we simply assume all the station pairs to have the same angular resolution, while the real distributions of stations of SKA1-low and LOFAR are scale dependent \citep{Haarlem2013, Dewdney2016}.

SKA1-low\footnote{The parameters used here are taken from the SKA website https://astronomers.skatelescope.org/} is designed to have 512 stations and expected to cover the frequency range 50--350 MHz (corresponding to a 21~cm signal at $3\lesssim z \lesssim 27$).
The sensitivity of each station is $S= 0.97\,\rm m^{2}/K$ at $\nu = 130.9\,\rm MHz$ ($z = 9.85$).
Its baselines will be arranged in a compact core with a diameter of 1~km and longer baselines up to 80~km.
An accurate noise calculation requires a realistic distribution of the antenna and the simulation of the uv-coverage.
Here we follow \cite{Ghara2017} and simply assume an angular resolution $\vartheta = 2 \,\rm arcmin$.
It should be kept in mind that a higher/lower resolution would lead to larger/smaller noise on the images and thus lower/higher signal-to-noise ratio (S/N), while a too low resolution would not be able to capture the details of the 21~cm signal around the QSO.

For LOFAR (see e.g. \citealt{ciardi2015}), we assume $\epsilon = 0.5$, $N_{\rm st} = 48$,  $T_{\rm sys} = [140 + 60(\nu/300 \,\rm MHz)^{-2.55}]\,\rm K$ and $A_{\rm eff} = {\rm min}(\frac{\lambda^2}{3}, 1.5626) \times 16 \times 24\,\rm m^{2}$ (24 tiles times 16 dipoles per tile for one core station).
For consistency and a better comparison, also in this case we adopt $\vartheta = 2 \,\rm arcmin$.

We assume that foreground contamination can be removed without any residual, thus we do not include the foreground noise  \citep{Geil2008}.
For both SKA1-low and LOFAR, we take an integration time $t_{\rm int} = 3000 \,\rm hours$ and a frequency band $B = 0.2 \,\rm MHz$ corresponding to $\Delta z = 0.0166$ and a comoving length of 3.8~Mpc at $z = 9.85$.
To mimic the real observations then, in the following discussion we only consider a slice of the simulation box centered around the QSO with a width corresponding to $0.2 \,\rm MHz$.
To produce mock maps with pixel size $2 \,\rm arcmin$, we additionally average the physical properties of the cells in bins of width corresponding to $2 \,\rm arcmin$ (equivalent to $5.6$  comoving Mpc at $z=9.85$).
The S/N along the angular radius can then be defined as S/N $= N_{\rm pixel}^{1/2} |\delta \overline{T}_{\rm 21cm}| \delta T_{\rm N}^{-1}$, where $\delta \overline{T}_{\rm 21cm}$ is the average signal in the pixels contained within a bin of width $\vartheta$, and $N_{\rm pixel}$ is the number of pixels in the same bin.

\begin{figure}
    \centering
    \includegraphics[width=0.99\linewidth]{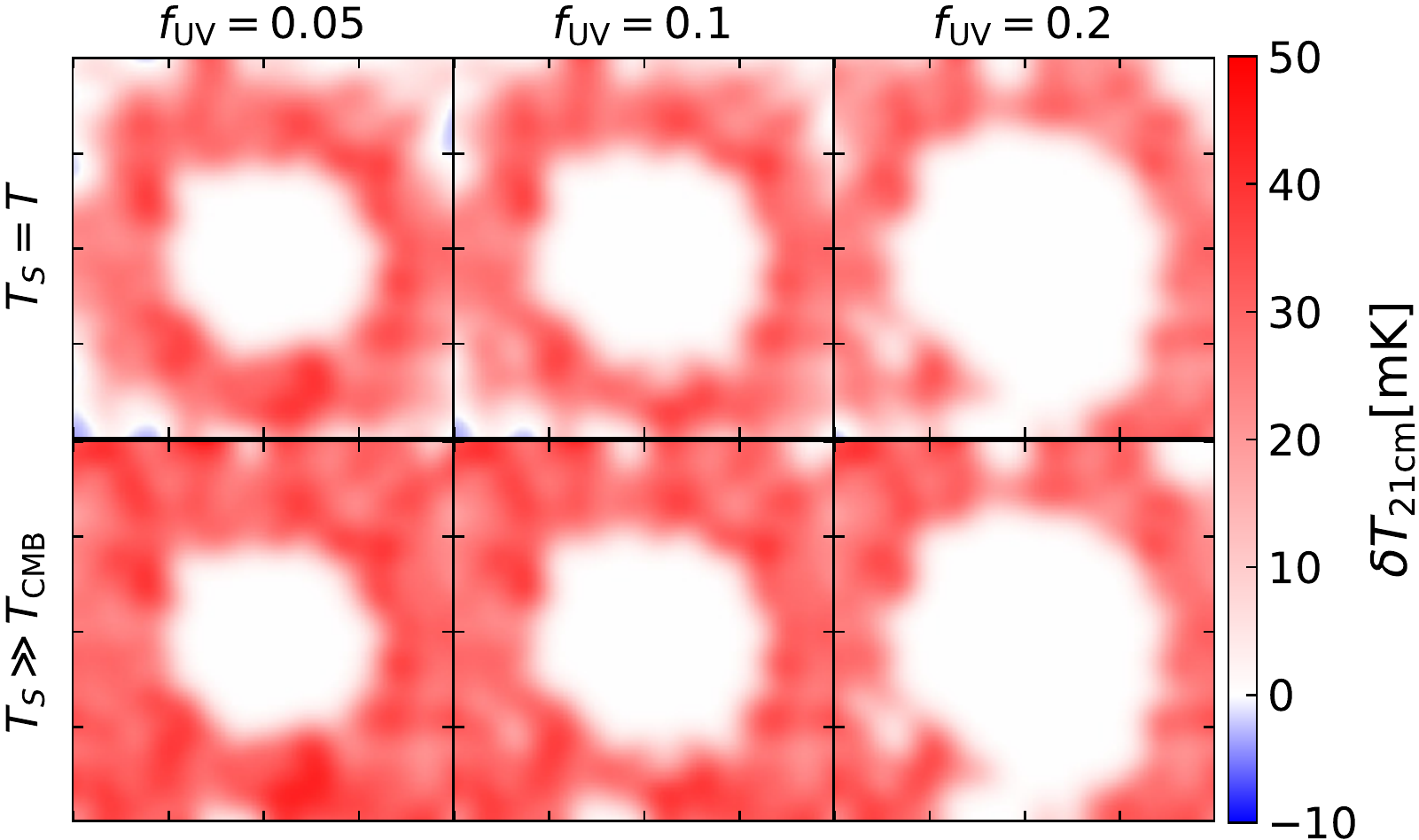}
    \caption{Central slices of 21~cm signal from the simulations with QSO in the case of $T_{S} = T$ (top) and $T_{S} \gg T_{\rm CMB}$ (bottom). From left to right, the columns refer to $f_{\rm UV} = 0.05$, 0.1 and 0.2. The maps refer to  $t_{\rm QSO}=10$~Myr (i.e. $z = 9.85$ and $\nu=130.9$~MHz) and have been obtained assuming an angular resolution of 2~arcmin. The thickness of the slices (3.8~Mpc) corresponds to a frequency band 0.2~MHz.
    }
    \label{fig:s_21cm_with_ts_ska}
\end{figure}
Before discussing the detectability of the signal, in Fig.~\ref{fig:s_21cm_with_ts_ska} we show a sample of mock observed 21~cm maps produced with an angular resolution of 2~arcmin. The maps refer to simulations including a QSO of $t_{\rm QSO}=10$~Myr, i.e. $z = 9.85$ and $\nu=130.9$~MHz, with three different $f_{\rm UV}$ values, assuming both $T_{S} = T$ (upper panels) and $T_{S} \gg T_{\rm CMB}$ (lower panels).
The maps are obtained by summing the central slices in the simulation contained within a comoving length of 3.8 Mpc (corresponding to $B = 0.2\,\rm MHz$), and then smoothing the resulting map with a gaussian kernel of 2~arcmin$/(2\sqrt{2ln2})$, i.e. 2.38 Mpc.
Compared to the theoretical maps in Fig.~\ref{fig:s_21cm_ts}, these mock observed maps show weaker fluctuations (as they are smoothed), while the central ionized bubble, as well as the emission from the heated IGM, are still clearly visible for both $T_{S} = T$ and $T_{S} \gg T_{\rm CMB}$. We should note that it will be difficult to capture these features with the 8~arcmin resolution expected for the SKA1-low core.

\begin{figure}
    \centering
    \includegraphics[width=0.99\linewidth]{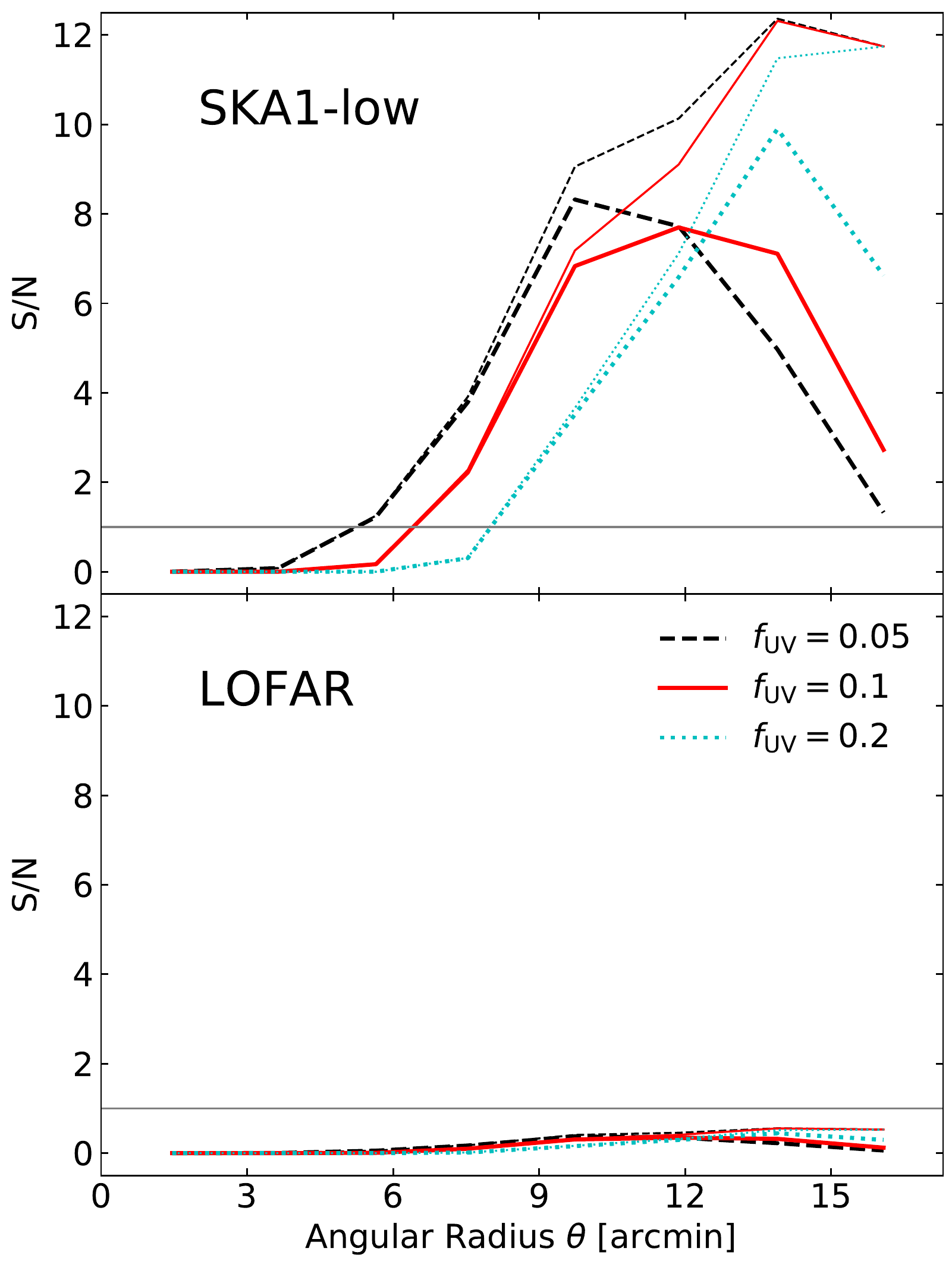}
    \caption{S/N expected from observations with SKA1-low (upper panel) and LOFAR (lower) in the simulations with a QSO turning on at $z=10$ and a lifetime of $t_{\rm QSO}=10$~Myr (i.e. $z = 9.85$ and $\nu=130.9$~MHz). The different lines refer to galactic emissivity of $f_{\rm UV} = 0.05$ (black dashed lines), 0.1 (red solid) and 0.2 (cyan dotted). The thick and thin lines refer to the case with $T_{S} = T$ and $T_{S} \gg T_{\rm CMB}$, respectively, while the horizontal line at S/N=1 is drawn to guide the eye. The S/N has been calculated assuming a bandwidth $B=0.2$~MHz, an integration time $t_{\rm int}=3000$~h and an angular resolution $\vartheta=2 \,\rm arcmin$.
    }
    \label{fig:s_21cm_mean_ang_sn}
\end{figure}
Fig.~\ref{fig:s_21cm_mean_ang_sn} shows the expected S/N of an observation of our reference QSO with $t_{\rm QSO}=10$~Myr both for SKA1-low and LOFAR.
While with LOFAR the S/N is always $<1$, for SKA1-low S/N values above 1 are obtained in a wide range of $\theta$, with a maximum of $\sim 8$ for $T_{S} = T$ and $>10$ for $T_{S} = T_{\rm CMB}$.
At distances larger than the ionized bubbles the S/N for $T_{S} \gg T_{\rm CMB}$ is much larger than that for $T_{S} = T$, due to the higher $\delta T_{\rm 21cm}$ (see Fig.~\ref{fig:s_21cm_mean_ang}).
While $f_{\rm UV}$ has little influence on the highest S/N attainable, larger luminosities shift the detection to higher values and a wider range of $\theta$.

\begin{figure}
    \centering
    \includegraphics[width=0.99\linewidth]{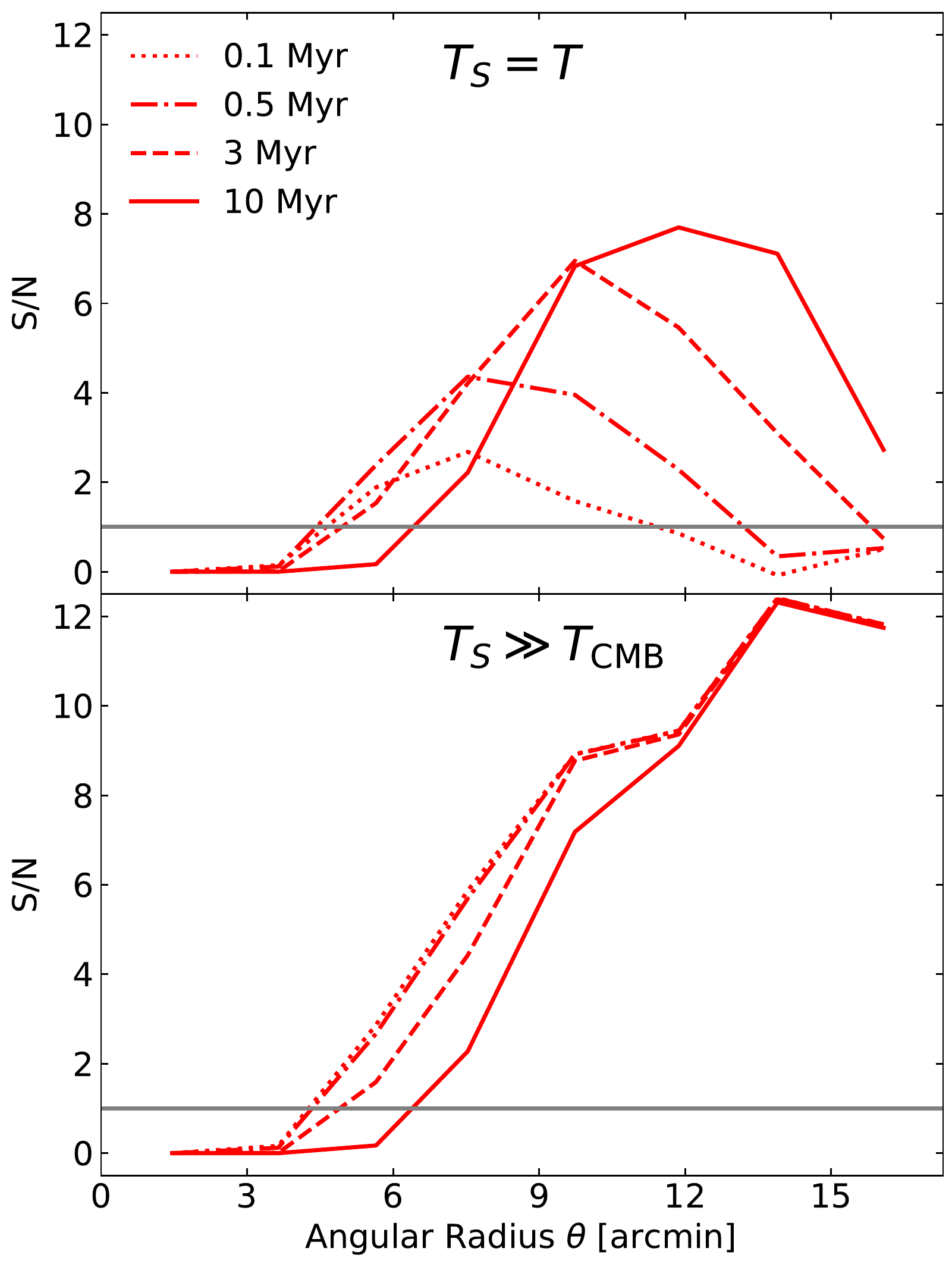}
    \caption{S/N expected from observations with SKA1-low in our fiducial model at $t_{\rm QSO}=0.1$~Myr (red dotted lines), 0.5~Myr (dash-dotted), 3~Myr (dashed) and 10~Myr (solid). The top and bottom panels refer to the case with $T_{S} = T$ and $T_{S} \gg T_{\rm CMB}$, respectively. The horizontal line at S/N=1 is drawn to guide the eye. The S/N has been calculated assuming an angular resolution $\vartheta=2 \,\rm arcmin$, a bandwidth $B=0.2$~MHz and an integration time $t_{\rm int}=3000$~h.}
    \label{fig:s_21cm_mean_ang_sn_ska}
\end{figure}
Fig.~\ref{fig:s_21cm_mean_ang_sn_ska} shows the expected SKA1-low S/N for our fiducial model at different $t_{\rm QSO}$.
With $T_{S} = T$, the highest S/N is $\sim 3$ at $t_{\rm QSO}=0.1$~Myr.
As $t_{\rm QSO}$ increases so does the S/N, and the signal from a  wider range of $\theta$ can be detected.
With $T_{S} \gg T_{\rm CMB}$, the 21 cm signal outside of the ionized bubble could be detected with a high S/N ($> 8$ at $\theta > 8 \,\rm arcmin$) for all the $t_{\rm QSO}$ considered here.

\begin{figure}
    \centering
    \includegraphics[width=0.99\linewidth]{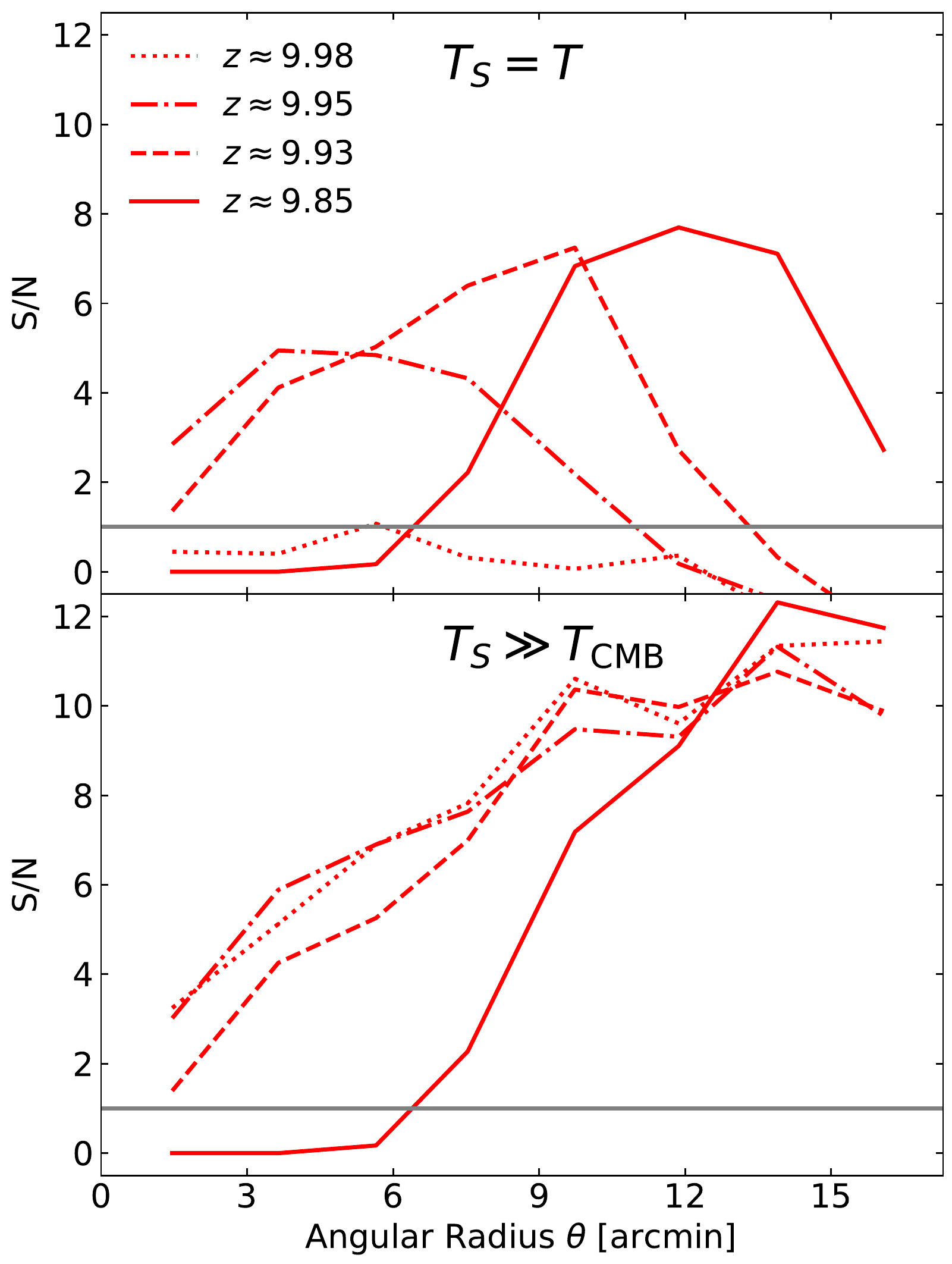}
    \caption{S/N expected from observations with SKA1-low of slices at $z = 9.98$ (dotted), $z = 9.95$ (dash-dotted) and $z = 9.93$ (dashed) and $z=9.85$ (solid) from the fiducial model with $t_{\rm QSO}=10$~Myr. The top and bottom panels refer to the case with $T_{S} = T$ and $T_{S} \gg T_{\rm CMB}$, respectively. The horizontal line at S/N=1 is drawn to guide the eye.}
    \label{fig:s_21cm_mean_ang_sn_ska_vz}
\end{figure}
Fig.~\ref{fig:s_21cm_mean_ang_sn_ska_vz} shows the expected SKA1-low S/N of slices perpendicular to the LOS at redshifts higher than the one of the QSO.
For $T_{S} = T$, while the S/N is $<1$ at $z= 9.98$, it can be $>5$ at the other redshifts. The range of $\theta$ over which the signal is detectable extends to larger values with decreasing redshift.
With $T_{S} \gg T_{\rm CMB}$ the S/N is always $>1$ in a wide range of $\theta$ at these redshifts.

We note that although LOFAR is not able to image the 21 cm signal around the QSO with $\vartheta=2\,\rm arcmin$, it is possible to observe the signal with a lower angular resolution and still give a good measurement of its evolution along the LOS.
For example, if we take $\vartheta = 12\,\rm arcmin$, which is enough to resolve the ionized bubble of the QSO in our fiducial simulation at $t_{\rm QSO} = 10\,\rm Myr$ (see e.g. Fig.~\ref{fig:s_21cm_mean_ang_los}), then the LOFAR noise is $= 11.4\,\rm mK$, giving a S/N at $z = 9.95$, 9.93 and 9.85 of 1.9, 1.7 and 0.06 if $T_{S} = T$, corresponding to an average $\delta T_{\rm 21cm} = 22$, 19.9 and 0.7~mK respectively. The S/N would be 2.5, 1.8 and 0.06 if $T_{S} = T_{\rm CMB}$, corresponding to an average $\delta T_{\rm 21cm} = 29.1$, 20.7 and 0.7~mK respectively.

\section{Discussion and Conclusion}
\label{sec:con}
Although the abundance of QSOs is expected to decline with increasing redshift \cite[e.g.][]{Fan2001, Khandai2015},
the possibility of detecting these very bright objects at $z\gtrsim 10$ with 21~cm line experiments is intriguing, as most hydrogen is expected to be still neutral at these high redshift, and thus the impact of heating from energetic sources such as QSOs, X-ray binaries (XRB) and shock heated interstellar medium (ISM) should be at  its maximum (see e.g. \citealt{Eide2018} and Eide et al. in prep), while it would be less relevant at lower $z$ when the Universe is highly ionized \citep{Datta2012}.
Additionally, while the detection of a bubble ionized by a QSO could be used to set constraints on QSO's characteristics such as its ionizing photon rate and/or lifetime (see e.g. \citealt{Datta2012}), resolving features associated to the QSO would become more challenging with decreasing redshift and merging of ionized regions  \citep{Furlanetto2005}.
However, also at lower $z$ a bright QSO might have significant effects on the surrounding 21 cm signal \citep{Alvarez2007, Geil2008}, e.g. by affecting the morphology of the HII regions they are born in \citep{Datta2012}.

We analyze the results of hydrodynamical and radiative transfer (RT) simulations \citep{Kakiichi2017} to study the 21 cm signal around a bright QSO at $z \sim 10$ and its detectability with SKA1-low and LOFAR.
The RT calculations include ionizing photons both from the QSO and the galaxies surrounding it.
As a reference, we also present the results of RT with only galaxies.
Besides, for each case we have three simulations with different emissivity efficiency of galaxies, $f_{\rm UV}$, to take into account the uncertainties of galactic luminosities and escape fraction during the epoch of reionization.
Finally, to include the uncertainties of gas heating at $z\sim 10$, we consider a case in which the spin temperature, $T_{S}$, has the same value of the gas temperature, i.e. $T_{S} = T$, and another one for which $T_{S} \gg T_{\rm CMB}$.

With the assumption that the radiation of the QSO is isotropic, our results of 21 cm signal show fairly spherically symmetric features.
However, the emission direction of the radiation though depends on the structure of the QSO's inner regions, leading typically to an anisotropic flux \citep{Elvis2000}.
This would change the morphology of the ionized bubbles, as well as the distribution of 21 cm emission signal. It should be noted, though, that such anisotropies would be partially washed out by the galaxies that also contribute abundant ionizing photons to the ionized region, and thus most of our conclusions should still be applicable.

Because of its strong radiation, the QSO can quickly heat its surroundings to a temperature $T \gg T_{\rm CMB}$, without fully ionizing the hydrogen component of the gas. As a consequence,
we find that for a QSO's lifetime of $t_{\rm QSO} = 10\,\rm Myr$ (corresponding to $z = 9.85$), the spherically averaged 21~cm brightness temperature, $\delta T_{\rm 21cm}$, in the case of $T_{S} = T$ displays a peak with a typical amplitude of $\sim 25 \,\rm mK$ and angular radius $\theta \sim 10\, \rm arcmin$, the former decreasing and the latter increasing with increasing $f_{\rm UV}$.
For $T_{S} \gg T_{\rm CMB}$ the 21~cm signal remains flat once $\delta T_{\rm 21cm}$ reaches its peak, whose amplitude and angular radius exhibit the same dependence on $f_{\rm UV}$ with the case of $T_{S} = T$.
As expected, the size of the ionized bubble and the radius of the 21~cm emission signal increase quickly with $t_{\rm QSO}$, a feature that would aid the determination of the age of the QSO \citep{Datta2012}.
Note that the measurement of both 21~cm images and rate of ionizing photons of galaxies and QSO are necessary to determine the lifetime of the QSO (see the discussions in \citealt{Datta2012}). Such an exercise will be additionally complicated by e.g. the anisotropic emission of QSOs and the uncertainties in the QSO and galaxy luminosity \citep{Euclid2019}, the latter increasing the error-bar in the determination of the QSO lifetime.
Although a detailed study of the impact of these quantities is beyond the scope of this paper, we believe that 3-D measurements of the 21~cm signal should give a rough picture of the anisotropic emission of the QSO and suggest whether the QSO is observed at an early or late evolutionary stage.

We then investigate the detectability of the signal around the QSO with SKA1-low and LOFAR.
At $z= 9.85$ (corresponding to $t_{\rm QSO} = 10\,\rm Myr$ and $\nu = 130.9\,\rm MHz$), with an integration time of 3000~h and an angular resolution of $2\,\rm arcmin$, SKA1-low could reach a signal-to-noise ratio of $\sim 8$ for $T_{S} = T$ and $>10$ for $T_{S} \gg T_{\rm CMB}$.
This is always lower than 1 in the case of LOFAR with the same integration time and angular resolution, while we find that with an angular resolution of $12\,\rm arcmin$, LOFAR is able to measure the 21 cm signal along the line of sight with a S/N $\sim 1-2$.
For $t_{\rm QSO} = 0.1-10\,\rm Myr$ the S/N for SKA1-low is expected to be larger than 1 in a wide range of angular radii.
Once the size of the ionized bubble and the 21 cm emission signal are resolved, one could set constraints on the lifetime of QSOs.
Note that with the matched filter technique discussed in \cite{Datta2012} a S/N higher than the one found in this work could be reached.
As $t_{\rm QSO}$ affects the 21 cm signal around the QSO in a way similar to $f_{\rm UV}$, an observation of the signal might not be able to break the degeneracy between these two parameters.
Considering the effect of finite light traveltime \cite[FLTT, ][]{Majumdar2011}, such degeneracy could be reduced by measuring the 3-D structure of the 21 cm signal, since the FLTT effect is expected to affect the 3-D image of signal around a QSO more than that around galaxies.
As an example, we have estimated expectation for our reference QSO at $z=9.85$, finding that SKA1-low could reach S/N $>1$ in a wide range of angular radii at redshifts as high as $z=9.98$.

\acknowledgments
This work is supported by innovation and entrepreneurial project of Guizhou province for high-level overseas talents (Grant no. (2019)02), National Natural Science Foundation of China (Grant No. 11847075, 11903010, 11565010 and U1731218), and Science and Technology Fund of Guizhou Province (Grant No. (2015)4015, (2016)4008, (2017)5726-37, (2018)5769-02).
The tools for bibliographic research are offered by the NASA Astrophysics Data Systems and by the JSTOR archive.


\bibliographystyle{aasjournal}
\bibliography{ref}

\end{document}